\documentclass[fp,twocolumn]{jpsj3}
\usepackage{txfonts}
\usepackage{bm}

\title{Performance of Zero-determinant Strategies in Repeated Games without Discounting}

\author{Masahiko Ueda$^1$\thanks{ueda@tmu.ac.jp}}
\inst{$^1$Department of Physics, Tokyo Metropolitan University, Hachioji, Tokyo 192-0397, Japan} 

\abst{Zero-determinant (ZD) strategies are a class of strategies in repeated games, which unilaterally control payoffs.
It has been shown that several ZD strategies promote cooperation in social dilemma games.
It has been widely believed that the performance of ZD strategies is expressed as ``unilaterally enforce linear relations between payoffs''.
However, in order to interpret properties of ZD strategies in repeated games without discounting, previous studies assumed that the limit of the Cesaro average of the probability distribution of the action profile exists.
Here, we explain the performance of ZD strategies without this assumption, which leads to a stronger result than previous ones.}

\begin{document}
\maketitle

\section{Introduction}
\label{sec:intro}
Social dilemma is a situation where rational behavior for individuals conflicts with the optimal outcome for society.
The prisoner's dilemma is a typical example of social dilemmas, where defection is rational for each individual whereas mutual cooperation improves the payoffs of both players \cite{FudTir1991}.
How cooperative behavior of autonomous agents can be realized is a long-standing question \cite{Now2006}.
One possible explanation for mutual cooperation is direct reciprocity, where repeated interaction among agents promotes mutual cooperation.
Originally, this explanation required rationality of agents \cite{FudTir1991}, which is far from human behavior.

The explanation for mutual cooperation based on direct reciprocity was modified by the appearance of zero-determinant (ZD) strategies \cite{PreDys2012,HCN2018}.
ZD strategies are a class of strategies in repeated games, which unilaterally enforce linear relations between payoffs.
In the repeated prisoner's dilemma game, several ZD strategies have been found.
The equalizer strategy unilaterally sets the opponent's payoff \cite{BNS1997}.
The extortionate strategy obtains payoff no less than that of the opponent \cite{PreDys2012}.
The generous strategy promotes mutual cooperation by adaptive learning of the opponent \cite{StePlo2013}.
Generally, if we can construct a positively-correlated ZD strategy \cite{CheZin2014}, which enforces a linear relation between the payoffs of two players with a positive slope, learning by the opponent increases the payoffs of both players.
The performance of ZD strategies in the context of evolution of cooperation has extensively been investigated \cite{HNS2013,AdaHin2013,StePlo2013,HNT2013,SzoPer2014,CWF2022}.
Furthermore, the utility of ZD strategies has also been studied in other games, such as the public goods game \cite{StePlo2013,PHRT2015}, the continuous donation game \cite{McAHau2016}, and the oligopoly game \cite{UYI2025}.
Mutual cooperation through ZD strategies is realized not by rationality but by the opponent's learning, and ZD strategies changed the problem of emergence of cooperation to the problem of control of other agents.

Usually, the performance of ZD strategies is expressed as ``unilaterally enforce linear relations between payoffs''.
Originally, ZD strategies were introduced in games without discounting about future payoffs, and properties of ZD strategies were investigated for the case in which all players use Markov strategies \cite{PreDys2012}.
After that, Akin proved that ZD strategies can control payoffs even if the co-players use non-Markov strategies \cite{Aki2016}, under the assumption that the limit of the Cesaro average of the probability distribution of the action profile exists.
Furthermore, it was also shown that, when there is discounting of future payoffs, this assumption is not necessary \cite{HTS2015}.
Recently, a necessary and sufficient condition for the existence of ZD strategies was specified for games without discounting \cite{Ued2022b}.
However, the performance of ZD strategies without the assumption about the limit has not been known when there is no discounting about future payoffs.

The purpose of this paper is to investigate properties of ZD strategies without the assumption about the existence of the limit of the Cesaro average.
We show that ZD strategies have stronger ability of payoff control than ``unilaterally enforce linear relations between payoffs''.
The performance of ZD strategies is described by some inequality.

\section{Preliminaries}
\label{sec:preliminaries}
We consider a repeated version of the normal-form game $G:=\left( \mathcal{N}, \left\{ A_j \right\}_{j\in \mathcal{N}}, \left\{ s_j \right\}_{j\in \mathcal{N}} \right)$, where $\mathcal{N}$ is the set of players, $A_j$ is the set of actions of player $j\in \mathcal{N}$, and $s_j: \prod_{k\in \mathcal{N}}A_k \rightarrow \mathbb{R}$ is the payoff function of player $j\in \mathcal{N}$ \cite{FudTir1991}.
We introduce the notation $\bm{a}:=\left( a_k \right)_{k\in \mathcal{N}}$ and $a_{-j}:=\left( a_k \right)_{k\neq j}$.
We assume that $A_j$ is a finite set for all $j\in \mathcal{N}$.
In repeated games, a strategy is a plan for choosing an action at each round when all previous action profiles are given.
As a special class of strategies, \emph{memory-one} strategies are ones using only information about the last round, that is, Markovian.
Typically, a memory-one strategy of player $j\in \mathcal{N}$ is described by $T_j\left( a_j | \bm{a}^\prime \right)$, which is the conditional probability to take $a_j\in A_j$ when the previous action profile is $\bm{a}^\prime=\left( a^\prime_k \right)_{k\in \mathcal{N}}$, with some initial distribution.

\emph{Zero-determinant} (ZD) strategies are a special class of memory-one strategies.
Concretely, a memory-one strategy $T_j$ of player $j$ is a ZD strategy controlling the quantity $B: \prod_{k\in \mathcal{N}}A_k \rightarrow \mathbb{R}$ if it satisfies
\begin{align}
 \sum_{a_j} c_{a_j} \left[ T_j \left( a_j | \bm{a}^\prime \right) - \delta_{a_j, a^\prime_j} \right] &= B\left( \bm{a}^\prime \right)
 \label{eq:ZDS}
\end{align}
with some coefficients $\left\{ c_{a_j} \right\}_{a_j \in A_j}$ and $B$ is not identically zero \cite{Ued2022b}, where $\delta_{a_j, a^\prime_j}$ is the Kronecker delta.
When we write the probability distribution of the action profile $\bm{a}$ at the $t$-th round by $P_t\left( \bm{a} \right)$, and introduce the the limit of the Cesaro average
\begin{align}
 P^*\left( \bm{a} \right) &:= \lim_{T\rightarrow \infty} \frac{1}{T} \sum_{t=1}^T P_t\left( \bm{a} \right),
\end{align}
a ZD strategy controlling $B$ unilaterally enforces \cite{Aki2016,Ued2022c}
\begin{align}
 \sum_{\bm{a}} B\left( \bm{a} \right) P^*\left( \bm{a} \right) &= 0.
 \label{eq:linear}
\end{align}
Therefore, when $B$ is related to the one-shot payoffs $\left\{ s_k \right\}_{k\in \mathcal{N}}$, a ZD strategy can control payoffs.
If there exists a stationary distribution $P^{(\mathrm{st})}\left( \bm{a} \right)=\lim_{t\rightarrow \infty} P_t\left( \bm{a} \right)$, $P^*$ coincides with $P^{(\mathrm{st})}$.
It should be noted that the payoff of player $k$ in repeated games without discounting is given by $\sum_{\bm{a}} s_k\left( \bm{a} \right) P^*\left( \bm{a} \right)$.
Originally, $B\left( \bm{a} \right)=\sum_{k\in \mathcal{N}} \alpha_k s_k \left( \bm{a} \right) + \alpha_0$ with some coefficients $\left\{ \alpha_k \right\}_{k\in \mathcal{N}\cup\{0\}}$ has been studied \cite{PreDys2012}, which corresponds to ``unilaterally enforce a linear relation between payoffs''.
However, this argument assumes the existence of $P^*$.

\section{Results}
\label{sec:results}
\subsection{Performance of ZD strategies}
Here, we investigate the performance of ZD strategies without assuming the existence of $P^*$, and prove a stronger result than Eq. (\ref{eq:linear}).
Below we write the action profile at the $t$-th round by $\bm{a}^{(t)}$.
Generally, the joint probability distribution of the action profiles $\left\{ \bm{a}^{(t^\prime)} \right\}_{t^\prime=1}^t$ in a repeated game is described as \cite{Ued2022c}
\begin{align}
 P\left( \left\{ \bm{a}^{(t^\prime)} \right\}_{t^\prime=1}^t \right) &= \left[ \prod_{t^\prime=2}^t \left[ \prod_{k\in \mathcal{N}} T_k^{(t^\prime)} \left( a_k^{(t^\prime)} | \left\{ \bm{a}^{(t^{\prime\prime})} \right\}_{t^{\prime\prime}=1}^{t^\prime-1} \right) \right] \right] \left[ \prod_{k\in \mathcal{N}} T_k^{(1)} \left( a_k^{(1)} \right) \right],
\end{align}
where $T_k^{(t^\prime)} \left( a_k^{(t^\prime)} | \left\{ \bm{a}^{(t^{\prime\prime})} \right\}_{t^{\prime\prime}=1}^{t^\prime-1} \right)$ is the player $k$'s conditional probability of taking $a_k^{(t^\prime)}$ in the $t^\prime$-th round given the history of the action profiles $\left\{ \bm{a}^{(t^{\prime\prime})} \right\}_{t^{\prime\prime}=1}^{t^\prime-1}$.
This joint probability distribution satisfies a recursion relation
\begin{align}
 P\left( \left\{ \bm{a}^{(t^\prime)} \right\}_{t^\prime=1}^{t+1} \right) &= \left[ \prod_{k\in \mathcal{N}} T_k^{(t+1)} \left( a_k^{(t+1)} | \left\{ \bm{a}^{(t^{\prime\prime})} \right\}_{t^{\prime\prime}=1}^{t} \right) \right] P\left( \left\{ \bm{a}^{(t^\prime)} \right\}_{t^\prime=1}^t \right).
 \label{eq:recursion_joint}
\end{align}
Here, we introduce the marginal distribution of $\bm{a}^{(t)}$:
\begin{align}
 P_t \left( \bm{a}^{(t)} \right) &:= \sum_{\left\{ \bm{a}^{(t^\prime)} \right\}_{t^\prime=1}^{t-1}} P\left( \left\{ \bm{a}^{(t^\prime)} \right\}_{t^\prime=1}^t \right).
\end{align}
When we calculate $\sum_{a_{-j}^{(t+1)}} \sum_{\left\{ \bm{a}^{(t^\prime)} \right\}_{t^\prime=1}^{t}}$ of Eq. (\ref{eq:recursion_joint}), we obtain
\begin{align}
 & \sum_{\bm{a}^\prime} \delta_{a_j^\prime, a_j^{(t+1)}} P_{t+1} \left( \bm{a}^\prime \right) \nonumber \\
 &= \sum_{\left\{ \bm{a}^{(t^\prime)} \right\}_{t^\prime=1}^{t}} T_j^{(t+1)} \left( a_j^{(t+1)} | \left\{ \bm{a}^{(t^{\prime\prime})} \right\}_{t^{\prime\prime}=1}^{t} \right) P\left( \left\{ \bm{a}^{(t^\prime)} \right\}_{t^\prime=1}^t \right).
\end{align}
If player $j$ uses a memory-one strategy, this is rewritten as
\begin{align}
 \sum_{\bm{a}^\prime} \delta_{a_j^\prime, a_j^{(t+1)}} P_{t+1} \left( \bm{a}^\prime \right) &= \sum_{\bm{a}^\prime} T_j \left( a_j^{(t+1)} | \bm{a}^\prime \right) P_t\left( \bm{a}^\prime \right).
\end{align}
By renaming $a_j^{(t+1)} \rightarrow a_j$, we obtain
\begin{align}
 \sum_{\bm{a}^\prime} \delta_{a_j, a^\prime_j} P_{t+1}\left( \bm{a}^\prime \right) &= \sum_{\bm{a}^\prime} T_j \left( a_j | \bm{a}^\prime \right) P_{t}\left( \bm{a}^\prime \right)
 \label{eq:before_akin}
\end{align}
regardless of the co-players' strategies.
It should be noted that the both sides describe the probability to take $a_j$ at the $(t+1)$-th round.
Suppose that $T_j$ is a ZD strategy (\ref{eq:ZDS}) controlling $B$.
By taking $\sum_{a_j} c_{a_j}$ of the both sides, we obtain
\begin{align}
 0 &= \sum_{a_j} c_{a_j} \sum_{\bm{a}^\prime} T_j \left( a_j | \bm{a}^\prime \right) P_{t}\left( \bm{a}^\prime \right) - \sum_{a_j} c_{a_j} \sum_{\bm{a}^\prime} \delta_{a_j, a^\prime_j} P_{t+1}\left( \bm{a}^\prime \right) \nonumber \\
 &= \sum_{\bm{a}^\prime} \left[ \sum_{a_j} c_{a_j} \delta_{a_j, a^\prime_j} + B\left( \bm{a}^\prime \right) \right] P_{t}\left( \bm{a}^\prime \right) - \sum_{\bm{a}^\prime} c_{a^\prime_j} P_{t+1}\left( \bm{a}^\prime \right) \nonumber \\
 &= \sum_{\bm{a}^\prime} c_{a^\prime_j} \left[ P_{t}\left( \bm{a}^\prime \right) - P_{t+1}\left( \bm{a}^\prime \right) \right] + \sum_{\bm{a}^\prime} B\left( \bm{a}^\prime \right) P_{t}\left( \bm{a}^\prime \right).
\end{align}
Then, by taking $\sum_{t=1}^T$ of the both sides with arbitrary $T\in \mathbb{N}$, we obtain
\begin{align}
 0 &= \sum_{t=1}^T \sum_{\bm{a}^\prime} c_{a^\prime_j} \left[ P_{t}\left( \bm{a}^\prime \right) - P_{t+1}\left( \bm{a}^\prime \right) \right] + \sum_{t=1}^T \sum_{\bm{a}^\prime} B\left( \bm{a}^\prime \right) P_{t}\left( \bm{a}^\prime \right) \nonumber \\
 &= \sum_{\bm{a}^\prime} c_{a^\prime_j} \left[ P_{1}\left( \bm{a}^\prime \right) - P_{T+1}\left( \bm{a}^\prime \right) \right] + \sum_{t=1}^T \sum_{\bm{a}^\prime} B\left( \bm{a}^\prime \right) P_{t}\left( \bm{a}^\prime \right).
\end{align}
This implies
\begin{align}
 \sum_{t=1}^T \sum_{\bm{a}^\prime} B\left( \bm{a}^\prime \right) P_{t}\left( \bm{a}^\prime \right) &= \sum_{\bm{a}^\prime} c_{a^\prime_j} \left[ P_{T+1}\left( \bm{a}^\prime \right) - P_{1}\left( \bm{a}^\prime \right) \right] \nonumber \\
 &\leq \max_{a_j} c_{a_j} - \min_{a_j} c_{a_j}
\end{align}
and
\begin{align}
 \sum_{t=1}^T \sum_{\bm{a}^\prime} B\left( \bm{a}^\prime \right) P_{t}\left( \bm{a}^\prime \right) &\geq \min_{a_j} c_{a_j} - \max_{a_j} c_{a_j}.
\end{align}
Thus, we finally obtain our main result
\begin{align}
 \left| \sum_{t=1}^T \sum_{\bm{a}^\prime} B\left( \bm{a}^\prime \right) P_{t}\left( \bm{a}^\prime \right) \right| &\leq \max_{a_j} c_{a_j} - \min_{a_j} c_{a_j}.
 \label{eq:linear_stronger}
\end{align}

Therefore, even if $P^*$ does not exist, ZD strategies can still control payoffs in the sense of Eq. (\ref{eq:linear_stronger}).
Note that the right-hand side does not depend on $T$, and Eq. (\ref{eq:linear_stronger}) means that total expected $B$ is bounded by some constant.
Particularly, if we consider $\lim_{T\rightarrow \infty} \frac{1}{T}$ of
\begin{align}
 \min_{a_j} c_{a_j} - \max_{a_j} c_{a_j} &\leq \sum_{t=1}^T \sum_{\bm{a}^\prime} B\left( \bm{a}^\prime \right) P_{t}\left( \bm{a}^\prime \right) \leq \max_{a_j} c_{a_j} - \min_{a_j} c_{a_j},
\end{align}
we obtain
\begin{align}
 \lim_{T\rightarrow \infty} \frac{1}{T} \sum_{t=1}^T \sum_{\bm{a}^\prime} B\left( \bm{a}^\prime \right) P_{t}\left( \bm{a}^\prime \right) &= 0.
 \label{eq:linear_stronger_limit}
\end{align}
That is, ZD strategies unilaterally enforce Eq. (\ref{eq:linear_stronger_limit}) even if $P^*$ does not exist.
When $P^*$ exists, Eq. (\ref{eq:linear_stronger_limit}) coincides with Eq. (\ref{eq:linear}).
Therefore, Eq. (\ref{eq:linear_stronger}) is stronger than Eq. (\ref{eq:linear}).

\subsection{Akin's lemma}
\label{subsec:akin}
Usually, properties (\ref{eq:linear}) of ZD strategies (\ref{eq:ZDS}) are derived by the Akin's lemma \cite{Aki2016,Ued2022c}:
\begin{align}
 \sum_{\bm{a}^\prime} \left[ T_j \left( a_j | \bm{a}^\prime \right) - \delta_{a_j, a^\prime_j} \right] P^*\left( \bm{a}^\prime \right) &= 0 \quad (\forall a_j).
 \label{eq:akin}
\end{align}
However, this lemma also requires the existence of $P^*$.
Here, we consider the extension of the Akin's lemma without assuming the existence of $P^*$.

According to Eq. (\ref{eq:before_akin}), we obtain
\begin{align}
  \sum_{\bm{a}^\prime} \left[ T_j \left( a_j | \bm{a}^\prime \right) - \delta_{a_j, a^\prime_j} \right] P_{t}\left( \bm{a}^\prime \right) &= \sum_{\bm{a}^\prime} \delta_{a_j, a^\prime_j} \left[ P_{t+1}\left( \bm{a}^\prime \right) - P_{t}\left( \bm{a}^\prime \right) \right]
\end{align}
for all $a_j$.
By taking $\sum_{t=1}^T$ of the both sides with arbitrary $T\in \mathbb{N}$, we obtain
\begin{align}
  & \sum_{t=1}^T \sum_{\bm{a}^\prime} \left[ T_j \left( a_j | \bm{a}^\prime \right) - \delta_{a_j, a^\prime_j} \right] P_{t}\left( \bm{a}^\prime \right) \nonumber \\
  &= \sum_{t=1}^T \sum_{\bm{a}^\prime} \delta_{a_j, a^\prime_j} \left[ P_{t+1}\left( \bm{a}^\prime \right) - P_{t}\left( \bm{a}^\prime \right) \right] \nonumber \\
  &= \sum_{\bm{a}^\prime} \delta_{a_j, a^\prime_j} \left[ P_{T+1}\left( \bm{a}^\prime \right) - P_{1}\left( \bm{a}^\prime \right) \right].
\end{align}
Therefore, we obtain
\begin{align}
  \left| \sum_{t=1}^T \sum_{\bm{a}^\prime} \left[ T_j \left( a_j | \bm{a}^\prime \right) - \delta_{a_j, a^\prime_j} \right] P_{t}\left( \bm{a}^\prime \right) \right| &\leq 1
  \label{eq:akin_stronger}
\end{align}
for arbitrary $a_j$.
This is a modified version of the Akin's lemma, which does not assume the existence of $P^*$.

Similarly as Eq. (\ref{eq:linear_stronger_limit}), if we consider $\lim_{T\rightarrow \infty} \frac{1}{T}$ of
\begin{align}
 -1 &\leq \sum_{t=1}^T \sum_{\bm{a}^\prime} \left[ T_j \left( a_j | \bm{a}^\prime \right) - \delta_{a_j, a^\prime_j} \right] P_{t}\left( \bm{a}^\prime \right) \leq 1,
\end{align}
we obtain
\begin{align}
 \lim_{T\rightarrow \infty} \frac{1}{T} \sum_{t=1}^T \sum_{\bm{a}^\prime} \left[ T_j \left( a_j | \bm{a}^\prime \right) - \delta_{a_j, a^\prime_j} \right] P_{t}\left( \bm{a}^\prime \right) &= 0.
 \label{eq:akin_stronger_limit}
\end{align}
This equality holds even if $P^*$ does not exist.
When $P^*$ exists, Eq. (\ref{eq:akin_stronger_limit}) coincides with Eq. (\ref{eq:akin}).

\subsection{Numerical simulation}
\label{subsec:numerical}
In order to check the validity of our theoretical result (\ref{eq:linear_stronger}), we perform a numerical simulation for a simple game.
Concretely, we consider the prisoner's dilemma game \cite{PreDys2012}, where $\mathcal{N} = \{ 1, 2 \}$ and $A_1=A_2=\{ C, D \}$.
The actions $C$ and $D$ correspond to cooperation and defection, respectively.
The payoffs are given by
\begin{align}
 \bm{s}_1 &:=  \left( s_1(C, C), s_1(C, D), s_1(D, C), s_1(D, D) \right)^\mathsf{T} \nonumber \\
  &= \left( R, S, T, P \right)^\mathsf{T} \nonumber \\
 \bm{s}_2 &:=  \left( s_2(C, C), s_2(C, D), s_2(D, C), s_2(D, D) \right)^\mathsf{T} \nonumber \\
  &= \left( R, T, S, P \right)^\mathsf{T},
\end{align}
where $T>R>P>S$ and $2R>T+S$ hold.

For this game, we consider the equalizer strategy of player $1$ \cite{BNS1997,PreDys2012}:
\begin{align}
 \bm{T}_1(C) &:=  \left( T_1(C|C, C), T_1(C|C, D), T_1(C|D, C), T_1(C|D, D) \right)^\mathsf{T} \nonumber \\
  &= \left( 1-\frac{R-r}{W}, 1-\frac{T-r}{W}, -\frac{S-r}{W}, -\frac{P-r}{W} \right)^\mathsf{T}
  \label{eq:equalizer_RPD}
\end{align}
with $r\in [P, R]$ and $W:=\max_{\bm{a}}\left| s_2(\bm{a})-r \right|>0$ \cite{Ued2022b}.
Usually, the equalizer strategy is described as a strategy which unilaterally enforces $\sum_{\bm{a}} s_2\left( \bm{a} \right) P^*\left( \bm{a} \right)=r$.
In other words, the equalizer strategy unilaterally sets the opponent's payoff in repeated games.
The equalizer strategy is a ZD strategy controlling $B=s_2-r$, because it satisfies Eq. (\ref{eq:ZDS}) as
\begin{align}
 \bm{T}_1(C) - (1, 1, 0, 0)^\mathsf{T} &= \left( -\frac{R-r}{W}, -\frac{T-r}{W}, -\frac{S-r}{W}, -\frac{P-r}{W} \right)^\mathsf{T} \nonumber \\
 &= -\frac{1}{W}\left[ \bm{s}_2 - r (1, 1, 1, 1)^\mathsf{T} \right].
\end{align}
That is, $c_C=-W$ and $c_D=0$ in the definition (\ref{eq:ZDS}).
For this situation, our bound (\ref{eq:linear_stronger}) is described as
\begin{align}
 \left| \sum_{t=1}^T \left[ \sum_{\bm{a}^\prime} s_2\left( \bm{a}^\prime \right) P_{t}\left( \bm{a}^\prime \right) - r \right] \right| &\leq W.
 \label{eq:linear_stronger_equalizer}
\end{align}

\begin{figure}[tbp]
\begin{center}
\includegraphics[clip, width=8.0cm]{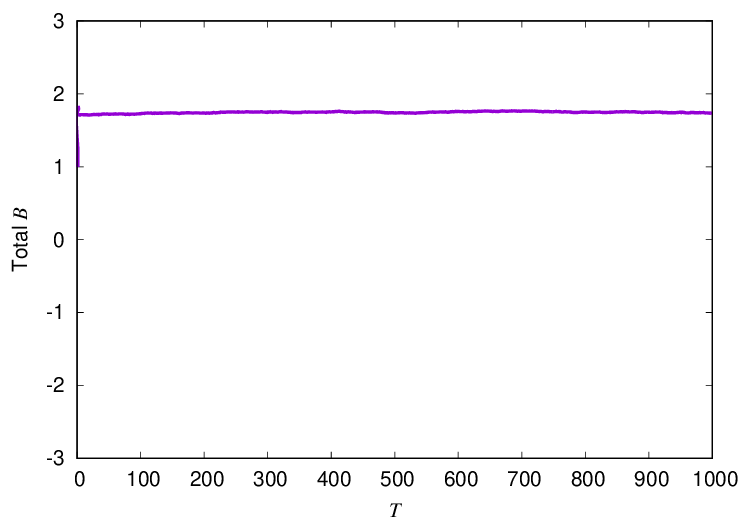}
\end{center}
\caption{Numerical verification of our theoretical result (\ref{eq:linear_stronger_equalizer}). The horizontal axis is $T$, and the vertical axis is $\sum_{t=1}^T \sum_{\bm{a}^\prime} B\left( \bm{a}^\prime \right) P_{t}\left( \bm{a}^\prime \right)$ with $B=s_2-r$, where the ensemble average is calculated from $10^6$ samples. The parameters are $(R, S, T, P)=(3, 0, 5, 1)$ and $r=(R+P)/2=2$. $W$ is calculated as $W=3$.}
\label{fig:withoutlimit_equalizer}
\end{figure}

Here, we set the parameters as $(R, S, T, P)=(3, 0, 5, 1)$ and $r=(R+P)/2=2$.
For this case, $W$ is calculated as $W=3$.
We consider the situation where player $1$ adopts the equalizer strategy (\ref{eq:equalizer_RPD}) and player $2$ takes a random action in $t\geq 2$.
The initial condition is set to $\bm{a}^{(1)}=(C, C)$.
The numerical result is displayed in Fig. \ref{fig:withoutlimit_equalizer}.
We find that the numerical result is consistent with our theory (\ref{eq:linear_stronger_equalizer}).

\section{Concluding remarks}
\label{sec:conclusion}
In this paper, we investigated the performance of ZD strategies without assuming the existence of $P^*$ in repeated games without discounting.
We proved that a ZD strategy controlling $B$ unilaterally enforces the inequality (\ref{eq:linear_stronger}).
Furthermore, this inequality leads to a ``linear relation'' (\ref{eq:linear_stronger_limit}).
Note that Eq. (\ref{eq:linear_stronger_limit}) is different from Eq. (\ref{eq:linear}), since the latter assumes the existence of $P^*$.
Therefore, even if $P^*$ does not exist, we can discuss the performance of ZD strategies.

For example, in the repeated prisoner's dilemma game, when player 1 adopts the Tit-for-Tat (TFT) strategy, which imitates the opponent's last action, and player 2 uses the strategy
\begin{align}
 a_2^{(t)} &=  \left\{
  \begin{array}{ll}
    C & \left( t\in \left[ 3^{2n}+1, 3^{2n+1} \right] \right) \\
    D & \left( t\in \left[ 3^{2n+1}+1, 3^{2n+2} \right] \right)
  \end{array}
  \right.
\end{align}
for each $n\geq 0$ with some initial condition, the limit $P^*$ does not exist.
Therefore, even though TFT is known to be a ZD strategy which unilaterally equalizes the payoffs of two players in the repeated prisoner's dilemma game \cite{PreDys2012}, one cannot interpret the property if $P^*$ does not exist.
In contrast, by using our framework, we can interpret the property of ZD strategies in terms of Eqs. (\ref{eq:linear_stronger}) and (\ref{eq:linear_stronger_limit}) with $B=s_1-s_2$.

In general two-player symmetric games, an inequality similar to Eq. (\ref{eq:linear_stronger}) was obtained for unbeatable imitation strategies \cite{DOS2012b,DOS2014}.
After that, the unbeatable Tit-for-Tat strategy was proved to be a ZD strategy which unilaterally equalizes the payoffs of two players \cite{Ued2022}.
Similar results were also obtained in two concurrent repeated games played in parallel \cite{Uedsub}.
Therefore, we can regard our results as an extension of the results in Refs. \cite{DOS2012b,DOS2014} from unbeatable imitation to arbitrary ZD strategies.

It should be noted that the problem considered in this paper is specific to repeated games without discounting.
When there is a discount factor $\delta\in [0, 1)$, $P^*$ is modified as \cite{HTS2015}
\begin{align}
 P^*\left( \bm{a} \right) &=  (1-\delta) \sum_{t=1}^\infty \delta^{t-1} P_t\left( \bm{a} \right),
\end{align}
which always exists.
Therefore, the results of this paper complete characterization of the performance of ZD strategies for all $\delta$.

We remark on the meaning of $c_{a_j}$ in Eq. (\ref{eq:ZDS}).
In construction of ZD strategies, the coefficients $\left\{ c_{a_j} \right\}_{a_j \in A_j}$ have been heuristically obtained \cite{Ued2022b}.
For potential games \cite{MonSha1996}, $c_{a_j}$ is related to potential functions \cite{Ued2022}.
Recently, a learning algorithm to obtain a ZD strategy together with $\left\{ c_{a_j} \right\}_{a_j \in A_j}$ was proposed \cite{MMHet2025}.
Since the coefficients $\left\{ c_{a_j} \right\}_{a_j \in A_j}$ determine the bound in the right-hand side of Eq. (\ref{eq:linear_stronger}), a ZD strategy with small $\Delta := \max_{a_j} c_{a_j} - \min_{a_j} c_{a_j}$ is more powerful.
How to obtain ZD strategies with small $\Delta$ should be investigated in future.

\begin{acknowledgment}
This study was supported by JSPS KAKENHI Grant Number JP26K21335, and JST PRESTO, Japan, Grant Number JPMJPR26K1.
\end{acknowledgment}

\bibliographystyle{jpsj}
\bibliography{ZDS}

\end{document}